\documentclass[twocolumn,english,prl,showpacs,superscriptaddress,longbibliography]{revtex4-1}
\usepackage{amsfonts}
\usepackage{amsmath}
\usepackage{amssymb}
\usepackage[colorlinks, citecolor=blue,linkcolor=red]{hyperref}
\usepackage{color}
\usepackage{hyperref}
\usepackage{textcomp}
\usepackage[rightcaption]{sidecap}
\usepackage{subfigure}
\usepackage[rightcaption]{sidecap}
\usepackage{graphicx}
\usepackage{lipsum}

\begin{document}
	\title{$\mathcal{PT}$ Symmetric Non-Hermitian Cavity Magnomechanics}
	\author{Yu Chengyong}
	\affiliation{Department of Physics, Zhejiang Normal University, Jinhua 321004, China.}
	\author{Kashif Ammar Yasir}
	\email{kayasir@zjnu.edu.cn}
	\affiliation{Department of Physics, Zhejiang Normal University, Jinhua 321004, China.}
	\setlength{\parskip}{0pt}
	\setlength{\belowcaptionskip}{-10pt}
	\begin{abstract}
	We design and explore PT-symmetric behavior of a hybrid non-Hermitian cavity magnomechanics consisting of a ferromagnetic YIG sphere driven by external magnetic field. Non-Hermicity is engineered by using a traveling field directly interacting with YIG. The external magnetic field excites collective mechanical modes of magnons, which later excites cavity mode leading to a coupling between cavity magnons and photons. The magnomechanical interaction of the system also excites phonon and couple them to the system. By computing eigenvalue spectrum, we demonstrate the occurrence of three-order exceptional point emerge with the increase of magnon-photon coupling at a specific incidence angle of traveling field. We illustrate the unique bi-broken and uni-protected PT-symmetry regions in eigenvalue spectrum unlike previously investigated non-Hermitian system, which can be tuned with gain and loss configuration by manipulating ratio between traveling field strength and magnon-photon coupling. Interestingly, protected PT-symmetry only exists on the axis of exceptional point. We further show that the PT-symmetry can only be govern at two angle of incident of traveling field. However, later, by performing stability analysis, we illustrate that the system is only stable at $\pi/2$ and, on all other angles, either the system is non-PT-symmetric or it is unstable. Furthermore, we govern the parametric stability conditions for the system and, by defining stablity parameter, illustrate the stable and unstable parametric regimes. Our finding not only discusses a new type of PT-symmetric system, but also could act as foundation to bring cavity magnomechanics to the subject of quantum information and process.
	\end{abstract}
	\date{\today}
	\maketitle

	In recent years, the cavity magnomechanics, containing both coherent magnon-phonon coupling and dissipative magnon-photon coupling, has sparked widespread research interest, particularly by using single-crystal yttrium iron garnet (YIG) spheres. Benefiting from high spin density and strong spin-spin exchange interactions \cite{PhysRev.73.155}, YIG sphere inside the cavity magnomechanics system can achieve strong\cite{PhysRevLett.111.127003,PhysRevLett.113.156401,zhang2015cavity} to even ultra-strong coupling\cite{PhysRevB.93.144420}. Direct coupling between magnons and vibration modes of YIG sphere can be achieved by magnetostrictive interaction, offering a novel platform for investigating strong-coupling cavity electrodynamics. Significant progress has been made in various related studies, including ground-state cooling\cite{ASJAD20233}, cavity-enhanced coherent scattering\cite{PhysRevLett.122.123602,PhysRevLett.122.123601}, tripartite entanglement among magnons, photons, and phonons\cite{PhysRevLett.121.203601}, as well as quantum steering\cite{PhysRevA.101.032120,PhysRevLett.114.060404}. These advancements underscore the immense developmental potential of such systems.
	
	The development of non-Hermitian physics has opened a new realm of investigation, giving rise to numerous novel areas of study and is the subject of increasing investigations. Traditionally, real eigenvalues were thought to be exclusive to Hermitian systems. However, research by C. M. Bender has shown that under Parity-time symmetry (PT-symmetry), non-Hermitian systems which violates the Hermitian condition $ \hat{H}\neq \hat{H}^\dagger $ can also possess real eigenvalues\cite{PhysRevLett.80.5243}. To date, exploration of the new physical phenomena brought about by PT symmetry in quantum optical systems has made notable progress, exemplified by advancements in areas such as Entanglement dynamics in anti-PT -symmetric systems\cite{PhysRevResearch.4.033022,PhysRevA.105.022404}, nonzero entropy under broken PT symmetry\cite{PhysRevResearch.3.013256}, and PT symmetry breaking enhanced cavity optomechanical magnetometry\cite{PhysRevA.102.023512}. 
	
	In addition, When the eigenvalue goes from complex to real, the system will experience a phase transition at an exceptional point (EP) where the eigenvalues coalesce together. When three or more eigenvalues coalesce, the system is said to exhibit a third-order EP or higher-order EP. Over the past decade, second-order EPs have been extensively studied both theoretically and experimentally, as seen in works such as quantum Rabi model\cite{PhysRevA.108.053712}, cavity optomechanical system\cite{PhysRevLett.113.053604,xu2016topological}, and cavity magnon-polaritons\cite{zhang2017observation}. The emergence of higher-order EPs introduces new physical effects to the system, often exhibiting superior characteristics compared to second-order EPs. These enhanced properties include stronger nonlinear responses\cite{dai2024non,PhysRevA.104.063508}, higher reconfigurability\cite{li2023exceptional,PhysRevLett.124.030401}, and other advanced features that make higher-order EPs particularly promising for various applications. However, it is still worthwhile to explore such PT-symmetric behavior in non-Hermitian cavity magnomechanics.
	
	In this paper, we introducing the PT-symmetry dynamics of a non-Hermitian quantum cavity magnomechanical system. The system interactions excite controlled magnons, photons, and phonons. The non-Hermitian behavior is achieved by using a traveling optical field directly exciting magnons and inducing gain in the system. By introducing the traveling field at specific angles and tuning the magnon-photon coupling rate, we govern the PT symmetric dynamics and demonstrate the existence of third-order EP in the system. Through the analysis of the gain and loss mechanisms within the subsystems, we establish the conditions for PT-symmetry. Further, we illustrate when non-Hermitian strength matches to the magnon-photon coupling rate, the system possesses protected PT-symmetry with real eigenvalues. We also show the novel occurrence of bi-broken and single protect PT-symmetry, which can be tuned with the traveling field strength. Furthermore, we perform the stability analysis and govern minimum stability condition for the system. We drive the stability parameter and govern the overall dynamics to visualize parametric stability. It reveals the requirements for the incident angle of the traveling field to ensure the stable existence of third-order EPs. Because the stable presence of higher-order EPs in a strongly coupled magnomechanical system holds significant potential for advancing research in quantum entanglement and quantum correlations, providing new insights and possibilities for the field of quantum information.
	
	\begin{figure}[tp]
		\includegraphics[width=8.5cm]{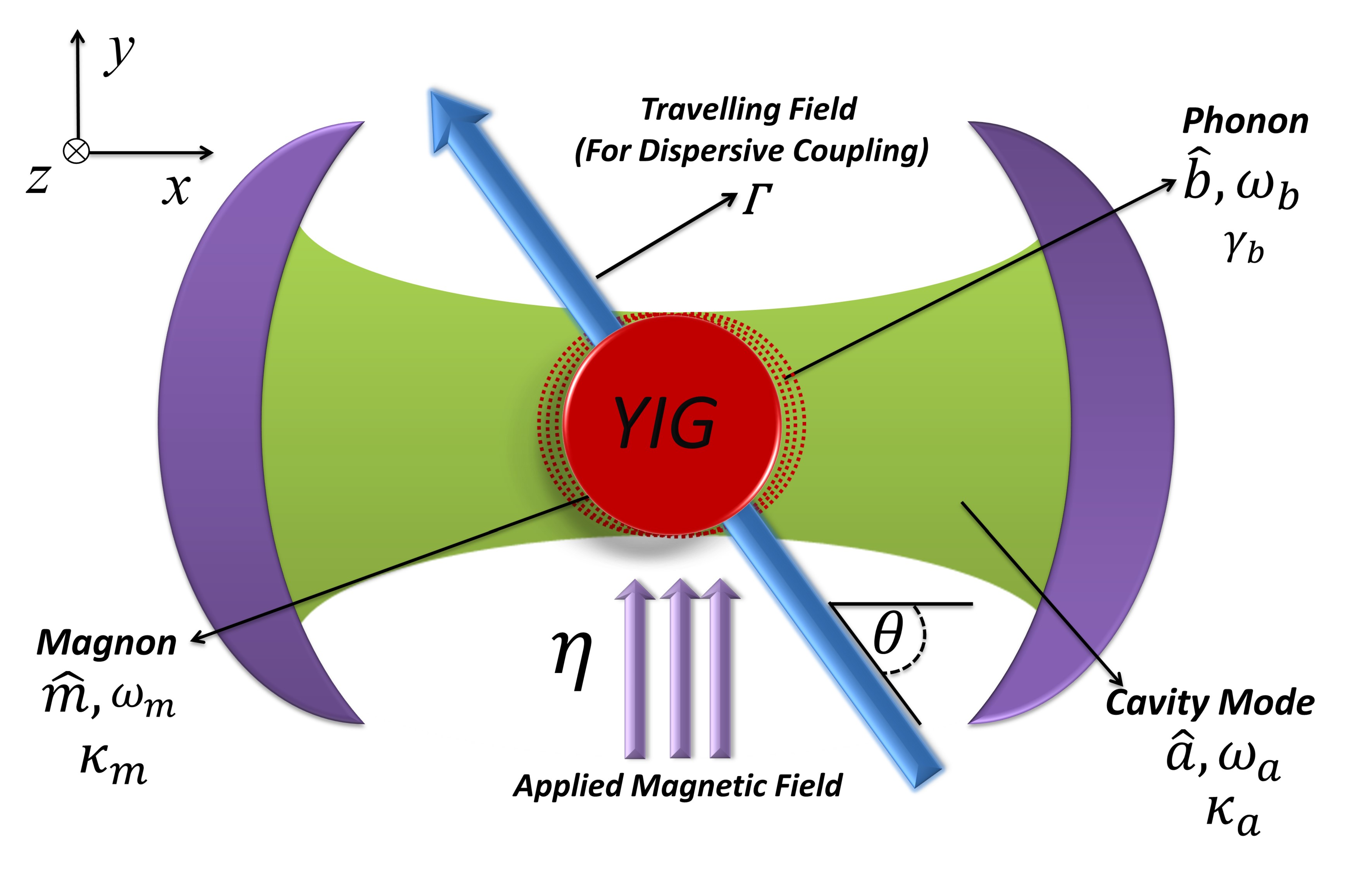}
		\caption{Schematic diagram of the non-hermitian cavity magnomechanical system. A YIG sphere is placed inside a microwave cavity driven by a strong magnetic field $\eta$ exciting magnons $\hat{m}$, which coupling with the cavity photons $\hat{a}$ (having frequency $\omega_a$) oscillating at $\omega_m$. Magnomechanical interactions also excite phonon $\hat{b}$ from YIG with frequency $\omega_b$. To make system non-Hermitain, a traveling field interacts with YIG at angle $ \theta $ and strength $\Gamma$.}
		\label{fig1}
	\end{figure}
	We consider a single-crysta YIG microsphere placed inside a microwave Fabry–Pérot cavity having length $L\approx12.5\times10^{-4}m$. The system is directly driven by a strong magnetic source with frequency $\omega_{0}=3.8\times2\pi\times10^{14}\mathrm{Hz}$ and power $P=0.0164\mathrm{mW}$ (along the $ y $-axis), which directly interacts and excites collective mechanical modes called as magnons from YIG. These modes oscillate with frequency $ \omega_m=2\pi\times10GHz $ with decay rate $ \kappa_m=2\pi\times1.5MHz $. The magnons gets later couple with scattering photons YIG via magnetic dipole interactions $ G_{ma}=4.3\times2\pi GHz, $. These scattering photons generate a strong cavity mode along $x$-axis because of the high-quality $Q$ factor of the cavity. Further, the magnetostrictive interactions, under the cavity mediated radiation pressure, excite phonons (mechanical modes) coupled to the magnons $ g_{mb}=2\pi\times2MHz $. The resonant frequencies of the microwave cavity mode and phonon mode are $ \omega_a=2\pi\times10GHz$ and $\omega_b=2\pi\times40GHz $, respectively. In order to discus experimentally feasibility of our work, we selected a particular set of parameters \cite{doi:10.1126/sciadv.1501286,yasir2015tunable,RevModPhys.86.1391,Fan_2023,PhysRevApplied.12.054031,yasir2016controlled,yasir2017,yasir2022,yasir2023,yasir2024,Luo:23}. However, these parametric values can be manipulated by the normalizing factor as we do in our study. 
	
	Under rotating-wave approximation, the total Hamiltonian of the system reads as,
	\begin{eqnarray}\label{eq1}
		\hat{H}&=&\hbar\omega_a\hat{a}^\dagger\hat{a}+\hbar\omega_m \hat{m}^\dagger\hat{m}+\hbar\omega_b\hat{b}\hat{b}^\dagger+g_{mb}\hat{m}^\dagger\hat{m}(\hat{b}+\hat{b}^\dagger)\nonumber\\
		&+&G_{ma}(\hat{a}+\hat{a}^\dagger)(\hat{m}+\hat{m}^\dagger)+i\eta(\hat{m}^\dagger e^{-i\omega_0 t}-\hat{m}e^{i\omega_0 t})\nonumber\\
		&-&i\Gamma e^{\delta t+i\theta}(\hat{a}+\hat{a}^\dagger)(\hat{m}+\hat{m}^\dagger),
	\end{eqnarray}
	where $ \hat{a}(\hat{a}^\dagger),\hat{m}(\hat{m}^\dagger)$ and $\hat{b}(\hat{b}^\dagger) $ are the annihilation (creation) operator of the cavity, magnon and phonon, respectively, with $ [\hat{O},\hat{O}^\dagger]=1(O=a,m,b) $. $\eta$ is the strength of applied magnetic field with relation $\vert\eta\vert=\sqrt{P\times\kappa/\hbar\omega_{0}}$. 	$G_{ma}=\sqrt{2}(\omega_{c}/L)x_{m}$\cite{PhysRevLett.113.083603,PhysRevLett.114.227201}  is the induced coupling among magnons and photons, where $x_{m}=\sqrt{\hbar/2m_m\omega_{m}}$ defines the zero point motion of magnon with stationary mass $m_m$ and frequency $\omega_{m}$. While the coupling between magnons and phonon can be defined as $G_b=\sqrt{2}(\omega_{c}/L)x_{b}$ having $x_{b}=\sqrt{\hbar/2m_b\omega_{b}}$ with mass $m_b$ and frequency $\omega_{b}$. $\Gamma=\mathcal{\alpha}\sqrt{(\hbar/\omega_m m_m)}$ represents the coupling strength of traveling field with the magnonic mode, where $\mathcal{\alpha}$ is the amplitude of the traveling field. $\delta$ and $\theta$ accounts for the frequency and incident angle traveling field which it makes with cavity axis. 
	
	By applying the rotating-wave approximation which $(\hat{a}+\hat{a}^\dagger)(\hat{m}+\hat{m}^\dagger)\rightarrow (\hat{a}^\dagger\hat{m}+\hat{a}\hat{m}^\dagger) $\cite{zhang2016cavity} and under the frame rotating at the drive frequency $ \omega_0 $, we drive Heisenberg equations of motion (or quantum Langevin equations) to incorporate the associated dissipation with subsystems, for details see section of supplementary martial. While driving Langevin equations, we also consider the traveling field time independent yielding in $e^{\delta t+i\theta}\rightarrow e^{i\theta}$. After this, we linearized quantum Langevin equations over their steady-states to incorporate the quantum fluctuations $ \hat{O}=\langle O \rangle+\delta \hat{O}$, where $O$ is a generic operator corresponding to associated subsystems. The linearized quantum Langevin equations will then read as, 
	\begin{equation}\label{eq4}
		\begin{aligned}
			\dot{\delta}\hat{a}=&-(i\Delta_a+\kappa_a)\delta\hat{a}-(iG_{ma}+\Gamma e^{i\theta})\delta\hat{m},\\
			\dot{\delta}\hat{m}=&-(i\Delta_m+\kappa_m)\delta\hat{m}-(iG_{ma}+\Gamma e^{i\theta})\delta\hat{a}-iG_{mb}\hat{m}\delta\hat{b},\\
			\dot{\delta}\hat{b}=&-(i\Delta_b+\gamma_b )\delta\hat{b}-iG_{mb}\delta\hat{m},
		\end{aligned}
	\end{equation}
	where $ G_{mb}=g_{mb}\langle m \rangle $ is the effective magnon-phonon coupling rate and $ \Delta_{a,m,b}=\omega_{a,m,b}-\omega_0$ are the detunings corresponding to the subsystems, respectively. The matrix representation of the effective Hamiltonian of the system as
	\begin{equation}
		\hat{H}_{eff}=\begin{pmatrix}
			\Delta_a+i\kappa_a & G_{ma}+i\Gamma e^{i\theta} & 0 \\
			G_{ma}+i\Gamma e^{i\theta} & \Delta_m+i\kappa_m & G_{mb} \\
			0 & G_{mb} & \Delta_b+i\gamma_b \\
		\end{pmatrix}
	\end{equation}
	
	\begin{figure}[tp]
		\includegraphics[width=8cm]{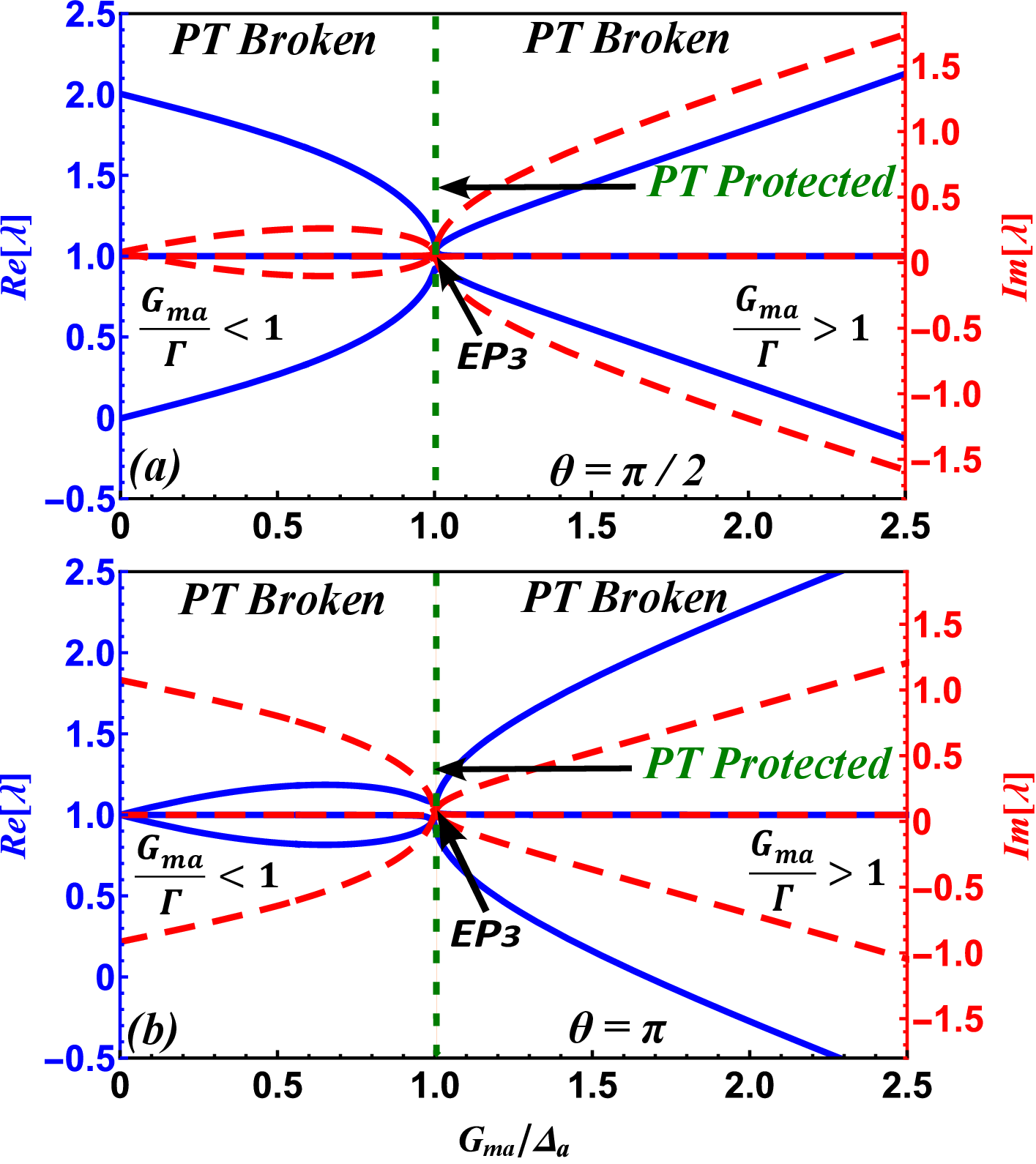}
		\caption{Eigenvalue spectrum as a function of normalized magnon-photon coupling $ G_{ma}/\Delta_a$, the blue solid line is the real part $Re[\lambda]$ while the red dashed line is for the imaginary part $Im[\lambda]$ of the eigenvalue. Here $ G_{ma}/\Delta_a=1 $ and $ \Gamma/\omega_b=1 $. (a) and (b) are for the incidence angles of $ \theta=pi/2 $ and $ \theta=\pi $ if traveling field, respectively. The other parameters that we choose are $ \kappa_a/\Delta_a=0.08,\kappa_m/\Delta_m=0.08, G_{mb}/\Delta_a=0.09, \gamma_b/\Delta_a=1/1000.$ } 
		\label{fig2}
	\end{figure}
	From the effective Hamiltonian, one can easily guess the occurrence of three eigenvalues for the hybrid magnomechanical system, as illustrated in Fig.\ref{fig2}. It reveals the relationship between the system’s eigenvalue (complex frequency) and dimensionless the magnon-photon coupling $ G_{ma}/\Delta_a=1 $ having fixed $ \Delta_a $. It can be observe that at $ G_{ma}/\Delta_a=1 $, the eigenvalue spectrum overlaps in such particular way that results in singularity intersection or third-order EP. Along this intersection, the system is operating in a protected PT symmetry domain and all three eigenvalues coexist in real space having no imaginary part. On the other hand, on both sides of the EP axis, the system is under broken PT symmetry region, which means that the eigenvalues are appeared to be complex and have both real and imaginary parts, as illustrated in Fig.\ref{fig2}(a) and \ref{fig2}(b). In both Figs.\ref{fig2}(a) and \ref{fig2}(b), one can also observe at $ G_{ma}=0 $, all eigenvalues also appear to have zero imaginary parts. But, as $ G_m{a}=0 $, it also means that the magnomechanical system is disintegrated and have no interaction resulting in trivial Hermitian state.
	
\begin{figure*}[htbp]
		\includegraphics[width=0.9\textwidth]{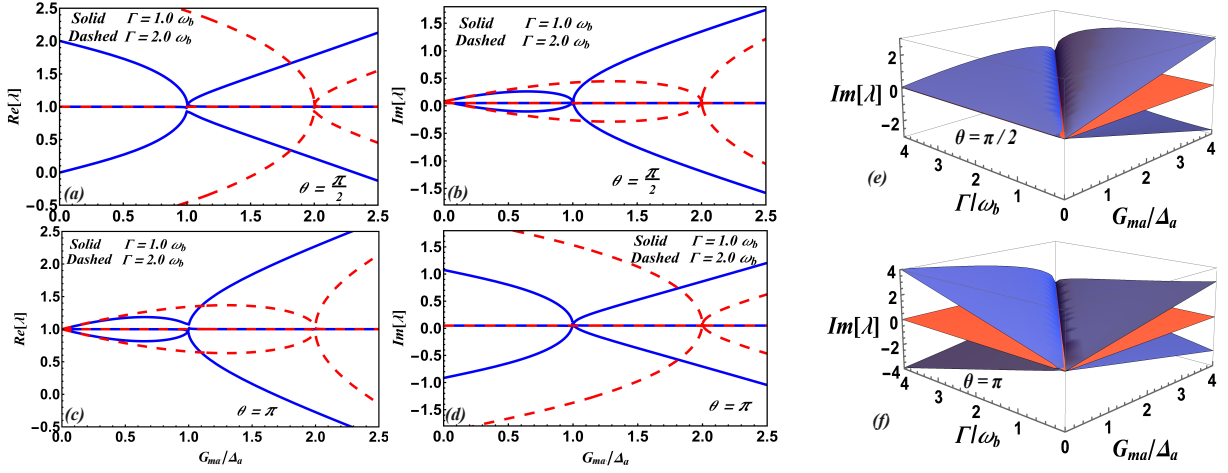}
		\caption{(a), (b), (c) and (d) are eigenvalue spectrum as a function of normalized $ G_{ma}/\Delta_a$. The incident angle $ \theta=pi/2 $ for (a) and (b), while $ \theta=\pi $ for (c) and (d), respectively. The solid line is $ \Gamma/\omega_b=1 $, the dashed line is $ \Gamma/\omega_b=2 $. (e) and (f) are 3D plot of the eigenvalue with respect to non-hermitian strength $ \Gamma/\omega_b $ and magnon-photon coupling $ G_{ma}/\Delta_a $, at angle $ \theta=pi/2 $ and $ \theta=\pi $, respectively. The other parameters are the same as in Fig.\ref{fig2}.} 
		\label{fig3}
	\end{figure*}

	As we know, in non-Hermitian optical systems, PT-symmetry breaking typically involves the imbalance between system's gain and loss configuration \cite{liu2024floquet,PhysRevLett.132.156901}, therefore, it is crucial to observe the influence of non-Hermitian parameter $\Gamma$ and its correspondence with the magnon-photon coupling of the system. It can be noted that the value of non-Hermitian parameter $\Gamma$ should be equal to the magnon-photon coupling, i.e. $G_{ma}=\Gamma=1$ at the axis of belonging to the EP. In other case, if $G_{a}\ne\Gamma$, then the system will possess broken PT, because eigenvalues will have both real and imaginary parts. At $G_{ma}/\Gamma\le 1$, the magnomechanical system appears to have lossy broken PT symmetry as the magnon-photon coupling is higher than the gain induced by the traveling field. While on other hand, at $G_{ma}/\Gamma\ge 1$, the system will operate with, sort-of, gain broken because the gain excited by the traveling field is not higher than the magnon-photon coupling. It will be further explained and verified by the Figs.\ref{fig3}(e) and \ref{fig3}(f).
	
	Further, the incident angle of the traveling field plays a crucial role in the non-Hermitian configuration of the system. From the Hamiltonian perspective, and according to Euler's formula, only if the angle of incidence is $ \pi/2, \pi, 3\pi/2, 2\pi $, the system can exist EPs. In our investigation, we found that the system only exhibits EPs when $ \theta=\pi/2 $ or $ \theta=\pi $, because at only these angles, the traveling field acts as a gain to the system. When this gain matches the magnon's loss, the system exhibits PT symmetry, naturally leading to the emergence of EPs. Because, at $\theta=\pi/2$ and $ \theta=\pi $, the non-Hermitian term in the Hamiltonian differs by an imaginary unit $ i $, yielding in opposite real and imaginary parts of the eigenvalue spectrum. This can be clearly observed by comparing Figs.\ref{fig2}(a) and \ref{fig2}(b). On the other hand, when $ \theta=3\pi/2 $ or $ \theta=2\pi $ the traveling field acts as a loss, providing photonic energy to the system, similar to the effect of a pump laser. Effects of such angles are discussed in detail in supplementary materials (.....). In such configuration, the system does not exhibit EPs because of the lossy environment. For angles other than these four specific values, the real and imaginary parts of the eigenvalues can merge under certain parameters configuration, but it will not contain EPs and the coupling strengths $ G_{ma} $ corresponding to these real and imaginary parts will be different, means they may not exist in equality relation corresponding to $ G_{ma} $ and $\Gamma$. Thus, EPs do not exist at incident angles other than $ \pi/2 $ and $ \pi $. One can note another interesting phenomenon that the real part $Re[\lambda]$ at $ \theta=\pi/2 $ is almost equal to the imaginary part at $ \theta=\pi $ and vise-versa. It can be easily understood by imagining complex variables getting orthogonality shifted over the complex plan. 
	
	Figs.\ref{fig3}(a-d) further delves into the impact of non-Hermitian strength on the system. As previously discussed, when the incident angle is $ \pi/2 $ and $ \pi $, system loss is positively correlated with non-Hermitian strength. When we increase the non-Hermitian strength $ \Gamma/\omega_b $ from $1$ to $2$, the EP also shifts from $ G_{ma}/\Delta_a=1 $ to $ G_{ma}/\Delta_a=2 $, maintaining the balance between system loss and gain. Therefore, we conclude that in this system, the PT-symmetry condition is achieved when the non-Hermitian strength equals the magnon-photon coupling rate, i.e.,$ \Gamma/\omega_b=G_{ma}/\Delta_a $. Alternatively saying, the increase in $ \Gamma/\omega_b $ shifts the interface between gain broken and lossy broken to higher magnon-photon coupling regions, providing more effective control over the position of EPs.
	
	In order to further enhance the understanding of the relation between $ G_{ma} $ and $\Gamma$, we plotted the eigenvalue spectrum versus these parameters, as illustrated in Figs.\ref{fig3}(e-f), where we plots the imaginary parts of the eigenvalues as functions of $ \Gamma/\omega_b $ and $ G_{ma}/\Delta_a $. One can note that the values or the parametric position illustrating the occurrence of EPs move diagonally between $ G_{ma} $ and $\Gamma$. Means, all of the eigenvalues coexist diagonally having zero value for all or the eigenvalues. Thus, the EPs exist at points where both $ G_{ma} $ and $\Gamma$ are equal with each other. It further proves our argument about the tunability of the non-Hermitian EPs. Both adjusting the traveling field to control the non-Hermitian strength and modifying the magnetic field to control the magnon-photon coupling rate offer significant operational flexibility.
	
	Parametric stability of such hybrid system is crucial because they operate in a multi and complex parametric configurations. To defined a particular set of parameters where system operates in a stable configuration, we use the Routh-Hurwitz stability criterion to develop minimum parametric stability conditions, for details see section two of supplementary materials. On these stability conditions, the system will remain stable. We strictly follow these conditions while choosing parameters in our numerical calculations. To further understand the stability behavior of the system, we drive a stability parameter from the Routh-Hurwitz stability conditions, reading as,
	\begin{equation}
		\begin{aligned}
			\mathcal{S}=&1+\frac{(G_{ma}-i\Gamma e^{i\theta})^2 (2\kappa_a \kappa_m \omega_b+\Delta_a (G_{mb}^2-2\omega_b \Delta_m)}{(\kappa_a^2+\Delta_a^2)(-G_{mb}^2\Delta_m}.... \\
			&....\frac{
				+(G_{ma}-i\Gamma e^{i\theta})^2 \omega_b)}{
				+\omega_b(\kappa_m^2+\Delta_m^2))},
		\end{aligned}	
	\end{equation}
	where $\Delta_m>0$ and $\Delta_a>0$ should be greater than zero while $\kappa_a+\kappa_m>\gamma_b/2$ in order to fulfill the stability conditions.
	\begin{figure*}[htp]
		\includegraphics[width=0.9\textwidth]{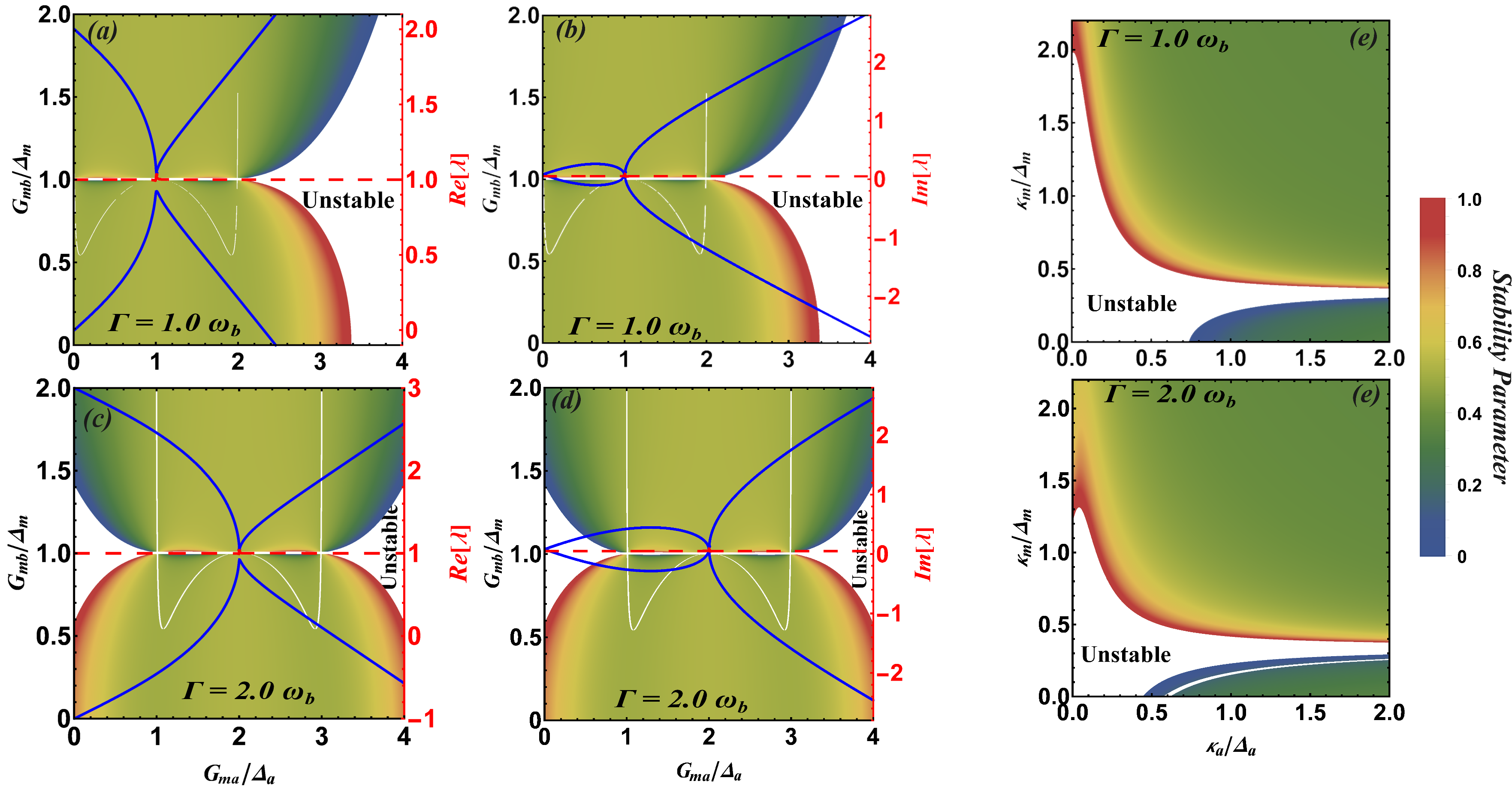}
		\caption{Stability parameter $\mathcal{S}$ as the function of magnon-photon coupling $ G_{ma} $ and effective magnon-phonon coupling $ G_{mb} $. (a) and (b) are for non-hermitian strength $ \Gamma/\omega_b=1.0 $, while (c) and (d) are for $ \Gamma/\omega_b=2.0 $, respectively. (e) and (f) illustrates $\mathcal{S}$ versus $ \kappa_a/\Delta_a $ and $ \kappa_m/\Delta_m $ at fixed $ G_{ma}/\Delta_a=\Gamma/\omega_b=1 $. The incident angle considered in plots is $ \theta=pi/2 $. The numerical parameters are as in Fig.\ref{fig2}} 
		\label{fig4}
	\end{figure*}
	
	From the stability parameter $\mathcal{S}$, one can graphically analysis the stability of the system, especially with respect to the incident angle of traveling field $\theta$, as illustrated in Fig.\ref{fig4}. The stability parameter $\mathcal{S}$ is plotted versus magnon-photon coupling $ G_{ma}/\Delta_a $ and the magnon-phonon coupling $ G_{mb}/\Delta_m $, with the traveling field incident angle set at $ \pi/2 $ and in comparison, with the PT-symmetric eigenvalues of the system. Here colored regions correspond to the stable regimes while white areas illustrate the instability of the system. It should be noted that the reason behind specifically choosing $ \theta=\pi/2 $ is that the system appears to be unstable at $ \theta=\pi $ and non-PT-symmetric on other angles. The reason is at when $ \theta=\pi $, the traveling field is parallel to the intracavity optical field and perpendicular to the magnetic field. This configuration functions similarly to adding a pump optical field along the x-axis, supplying photons and energy to the system, thereby driving it into an unstable state.
	
	Centering around the EP, the system exhibits two symmetric stability regions on either side of $ G_{mb}/\Delta_m=1 $, displaying strength vise opposite stability trends with increase in $ G_{ma}/\Delta_a $, as can be seen in Figs.\ref{fig4}(a,d). In regions with a lower magnon-phonon coupling rate, the system is more stable. When moving away from the EP, the system reaches maximum stability in regions of weak magnon-phonon coupling rate and minimum stability in regions of strong magnon-phonon coupling rate. The comparison here between stability and eigenvalues is only versus the magnon-photon coupling and other axis is independent. One can note that the system only appears to be unstable at higher values of $ G_{ma}/\Delta_a$ on both higher and lower regions of $ G_{mb}/\Delta_m$. 
	
	Further, the stability of the system crucially depends on the position of EP (or in other words, depends on the relation between non-Hermitian parameter and magnon-photon coupling), which is saturated between the center of stability regions. Therefore, any change in non-Hermitian parameter $\Gamma$ results in a shift in entire stability region similarly as they appear in eigenvalue spectrum, where EP shifts towards higher values of $ G_{ma}/\Delta_a$, as can be seen in Figs.\ref{fig4}(c,d). But it is interesting to note that similar unstable region, which was appearing at higher coupling rates, is now also appearing at lower coupling rates. It means that the collective system is now possessing bi-unstable parametric characteristics. 
	
	This analysis highlights the intricate balance of gain and loss in non-Hermitian systems and underscores the tunability of stability through both the magnon-photon coupling rate and the magnon-phonon coupling rate. Such tunability offers valuable insights for designing stable non-Hermitian systems and exploring their potential applications in quantum information science.
	
	In addition, we explored the impact of dissipations on system stability. The results reveal two clear boundaries between stability regions corresponding to higher and lower values, as shown in Figs.\ref{fig4}(e,f). The system tends to be more stable with higher magnon dissipation $\kappa_m$, while the dissipation of the optical field significantly determines the threshold for system stability. In conditions of low dissipation, it is challenging for the system to remain stable. However, non-Hermitian strength can partially compensate for this. Comparing Fig.\ref{fig4}(e) and \ref{fig4}(f), a stronger non-Hermitian strength allows the system to achieve stability under lower optical field dissipation conditions.

	In conclusion, we investigated a Cavity Magnomechanical system incorporating a single YIG sphere and studied non-Hermitian effects through a traveling field. We explored the PT-symmetry characteristics of the system and identified the conditions for PT-symmetry and EP, specifically when the non-Hermitian strength matches the magnon-photon coupling. Through the analysis of the eigenvalue spectrum, we confirmed the presence of a third-order EP in the system dividing spectrum into bi-broken PT regions (two broken and one unbroken), making our work unique from previous investigations. We also studied the stability conditions of the system, finding that stability regions exist around the EP. A smaller magnon-phonon coupling rate enhances system stability, whereas excessive cavity field dissipation leads to instability, however it can be tuned with non-Hermitian factor. Our research demonstrates that non-Hermitian effects can induce higher-order EPs in a Cavity Magnomechanical system, providing deeper insights into the applications and understanding of non-Hermitian physics.

	This work was supported by Research Fund for International Young Scientists by NSFC under grant No. KYZ04Y22050, Zhejiang Normal University research funding under grant No. ZC304021914 and Zhejiang province postdoctoral research project under grant number ZC304021952.

\end{document}


\title{Supplemental Material : $\mathcal{PT}$ Symmetric Non-Hermitian Cavity Magnomechanics}
	\author{Yu Chengyong}
\affiliation{Department of Physics, Zhejiang Normal University, Jinhua 321004, China.}
\author{Kashif Ammar Yasir}
\email{kayasir@zjnu.edu.cn}
\affiliation{Department of Physics, Zhejiang Normal University, Jinhua 321004, China.}

\date{\today}
\maketitle

\section{Heisenberg Equations of Motion}
By considering the rotating-wave approximation $(\hat{a}+\hat{a}^\dagger)(\hat{m}+\hat{m}^\dagger)\rightarrow (\hat{a}^\dagger\hat{m}+\hat{a}\hat{m}^\dagger)$ for the total Hamiltonian mentioned in main text \cite{zhang2016cavity} and applying the frame rotating at the drive frequency $ \omega_0 $, the total system Hamiltonian will read as,
\begin{equation}\begin{split}
		\hat{H}&= \Delta _{c}\hat{a}^{\dagger}\hat{a}+ \Delta _{m}\hat{m}^{\dagger}\hat{m}+ \Delta_{b}\hat{b}^{\dagger }\hat{b}\\
		&+(G_{ma}-\mathrm{i}\Gamma \mathrm{e}^{\mathrm{i}\theta  } )(\hat{a}\hat{m}^{\dagger }+\hat{a}^{\dagger }\hat{m}) +g_{mb}\hat{m}^{\dagger }\hat{m}(\hat{b}^{\dagger}+\hat{b}) \\
		&+\mathrm{i}\eta(\hat{m}^{\dagger} -\hat{m} ),
	\end{split} \label{H_tot}
\end{equation}
where $ \Delta_a=\omega_a-\omega_0, \Delta_m=\omega_m-\omega_0, \Delta_b=\omega_b-\omega_0 $ being the detuning of the microwave cavity mode, magnon mode, and phononic mode, respectively. After this, we drive quantum Langevin equations to incorporate the associated dissipation with subsystems and govern the time dynamics in the form of equations of motion, reading as, 
\begin{equation}\label{eq2}
	\begin{aligned}
		\dot{\hat{a}}=&-(i\Delta_a+\kappa_a)\hat{a}-(iG_{ma}+\Gamma e^{i\theta})\hat{m},\\
		\dot{\hat{m}}=&-(i\Delta_m+\kappa_m)\hat{m}-(iG_{ma}+\Gamma e^{i\theta})\hat{a}-ig_{mb}\hat{m}(\hat{b}+\hat{b}^\dagger),\\
		&+\eta\\
		\dot{\hat{b}}=&-(i\Delta_b+\gamma_b)\hat{b}-ig_{mb}\hat{m}\hat{m}^\dagger,
	\end{aligned}
\end{equation}

The steady-state mean values can be obtain from above equation by simply putting time-derivative equal to zero and solving them for individual subsystems, given as,
\begin{equation}
	\begin{aligned}
		\langle a \rangle=&-\frac{iG_{ma}+\Gamma e^{i\theta}}{i\Delta_a+\kappa_a}\langle m \rangle,\\
		\langle m \rangle=&\frac{\eta-(iG_{ma}+\Gamma e^{i\theta})\langle a \rangle}{i\Delta_m+\kappa_m+g_{mb}(b_s+b_s^*)},\\
		\langle b \rangle=&-\frac{ig_{mb}|\langle m \rangle|^2}{i\Delta_b+\gamma_b },
	\end{aligned}
\end{equation}

Since the magnon and cavity modes are strongly driven, resulting in large amplitude $ |\langle m\rangle|\gg1,|\langle a \rangle|\gg1 $, therefore, each Heisenberg operator can rewritten as a sum of its steady-state mean value and its corresponding quantum fluctuation $ \hat{O}=\langle O \rangle+\delta \hat{O} (O=a,m,b) $, as mentioned in the main text of the paper.
\\

\section{Routh-Hurwitz Stability Conditions}
By redefining associated subsystems in the form of position and momentum quadrature (i.e. $ \delta \hat{X}=(\delta \hat{a}+\delta\hat{a}^\dagger)/\sqrt{2} $, $ \delta \hat{Y}=i(\delta \hat{a}^\dagger-\delta\hat{a})/\sqrt{2} $, $ \delta \hat{x}=(\delta \hat{m}+\delta\hat{m}^\dagger)/\sqrt{2} $, $ \delta \hat{y}=i(\delta \hat{m}^\dagger-\delta\hat{m})/\sqrt{2}$, $ \delta \hat{q}=(\delta \hat{b}+\delta\hat{b}^\dagger)/\sqrt{2} $, $ \delta \hat{p}=i(\delta \hat{b}^\dagger-\delta\hat{b})/\sqrt{2}$), we rewrite linearized quantum Langevin equations in the following form,
\begin{equation}
	\dot{u}(t)=\mathcal{A} u(t),
\end{equation}
where $  u(t)=[\delta \hat{X}(t),\delta \hat{Y}(t),\delta \hat{x}(t),\delta \hat{y}(t), \delta q(t),\delta p(t)]^T $, and $\mathcal{A}$ is the drift or kernel matrix given as,
\begin{widetext}
	\begin{equation}
	\mathcal{A}=\begin{pmatrix}
			-\kappa_a  &  \Delta_a  & 0 &  g_{ma}-i\Gamma e^{i\theta}  & 0 & 0 \\
			\Delta_a  & - \kappa_a  & -( g_{ma}-i\Gamma e^{i\theta} ) & 0 & 0 & 0 \\
			0 &  g_{ma}-i\Gamma e^{i\theta}  & - \kappa_m  & \tilde{\Delta}_m & 0 & 0 \\
			-( g_{ma}-i\Gamma e^{i\theta} ) & 0 & -\tilde{\Delta}_m & - \kappa_m  &  -G_{mb}  & 0\\
			0 & 0 & 0 & 0 & 0 &  \omega_b \\
			0 & 0 &  -G_{mb}  & 0 &  -\omega_b  &  -\gamma_b 
		\end{pmatrix}
	\end{equation}
\end{widetext}
\begin{figure}[hp]
	\includegraphics[width=0.5\textwidth]{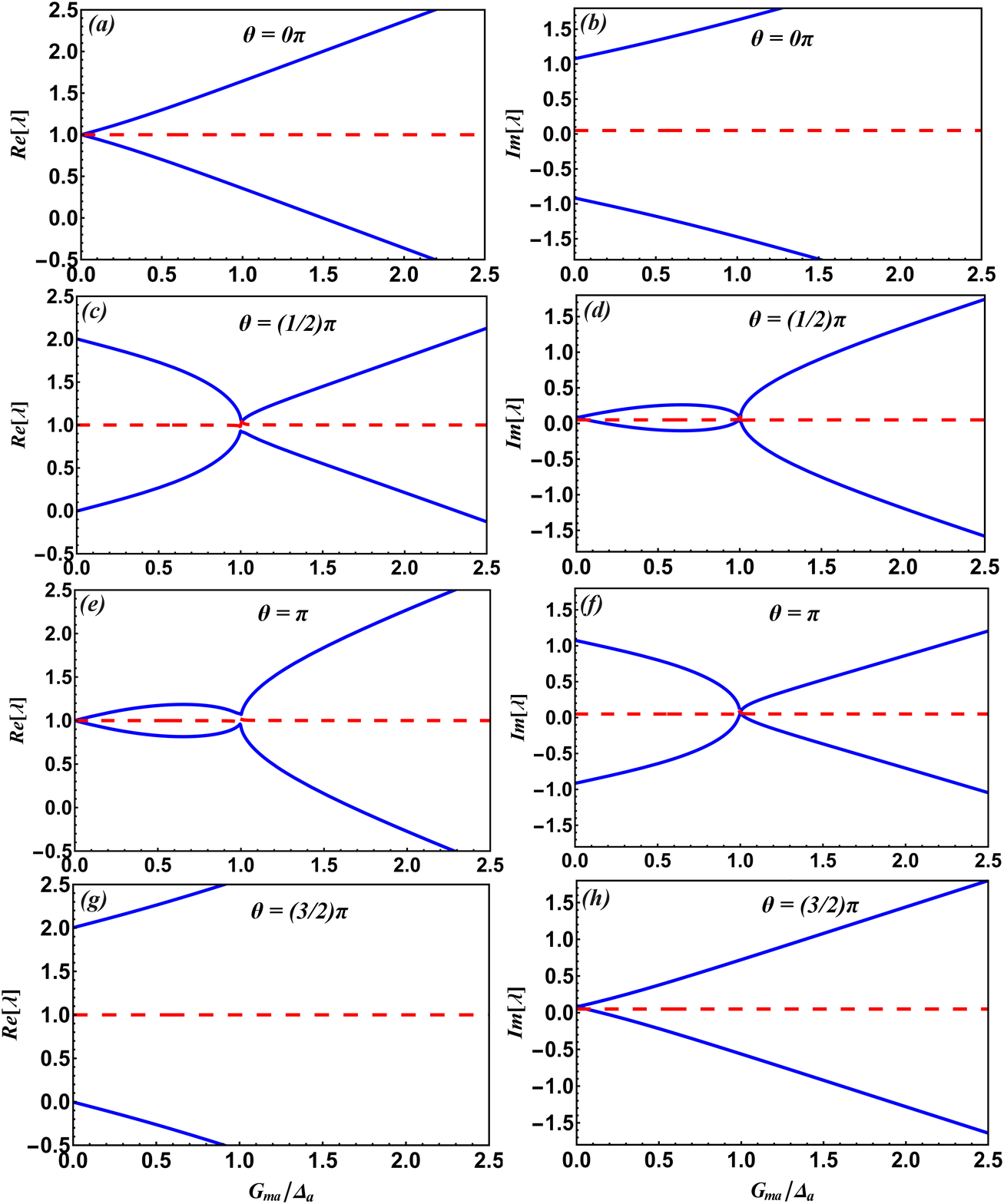}
	\caption{Eigenvalue spectrum contained both real $Re[\lambda]$ (left column) and imaginary part $Im[\lambda]]$ (right column) as a function of normalized magnon-photon coupling $ G_{ma}/\Delta_a$ at fixed traveling field strength $ \Gamma/\omega_b=1 $ and corresponding incident angles of traveling field are mentioned in the figures. The other parameters used in numerical calculations are same as in Fig.2 of main text.} 
	\label{figs1}
\end{figure}
Later, we apply Routh-Hurwitz stability criteria on the above matrix $\mathcal{A}$ in order to govern the stability conditions for the magnomechanical system \cite{Ref48,Ref29,Ref49}. Routh-Hurwitz stability criteria states that if any root of the characteristic polynomial of matrix $\mathcal{A}$ is on the left-half plan then the system will be unstable. By adopting and following that mechanism, we developed certain parametric stability conditions, given as,  
	\begin{equation}
	\begin{aligned}
		&(G_{ma}-i\Gamma e^{i\theta})^2 (2\kappa_a \kappa_m \omega_b + \Delta_a (G_{mb}^2-2\omega_b \Delta_m)\\
		&+(G_{ma}-i\Gamma e^{i\theta})^2 \omega_b)+(\kappa_a^2+\Delta_a^2)(-G_{mb}^2\Delta_m\\
		&+\omega_b(\kappa_m^2+\Delta_m^2))>0,\\
		&\kappa_a+\kappa_m>\frac{\gamma_b}{2},\\
		&\Delta_m>0,\\
		&\Delta_a>0.
	\end{aligned}
\end{equation}
We strictly follow these conditions while computing results presented and discussed in main text.

\section{Eigenvalue Spectrum and $\theta$ of Traveling Field}
In our model, the incident angle of the traveling field determines whether it acts as a gain or loss in the system. We investigate four incident angles where EPs could potentially exist: $ 0\pi, \pi/2, \pi,3 $ and $ \pi, $. When the incident angles are $ 0\pi $ and $ 3\pi/2 $, the traveling field acts as a loss, providing energy to the system. In this cases, no energy level merging occurs, and the system does not exhibit EPs. In other words, the eigenvalues of the system do not overlap to form an intersection for EP. This is clearly shown in Figs.\ref{figs1}(a), \ref{figs1}(b), \ref{figs1}(g) and \ref{figs1}(h). For specific photon-magnon couplings, the imaginary parts of the system's eigenenergies can be zero, but the real parts of the eigenenergies remain distinct, illustrated the absence of EPs. However, at other angle, i.e. $\pi/2$ and $\pi$, the eigenvalue spectrum contain EP and it is discussed detail in main text of the manuscript.

\section{Influence of Magnon-Phonon Coupling}
\begin{figure}[tp]
	\includegraphics[width=0.5\textwidth]{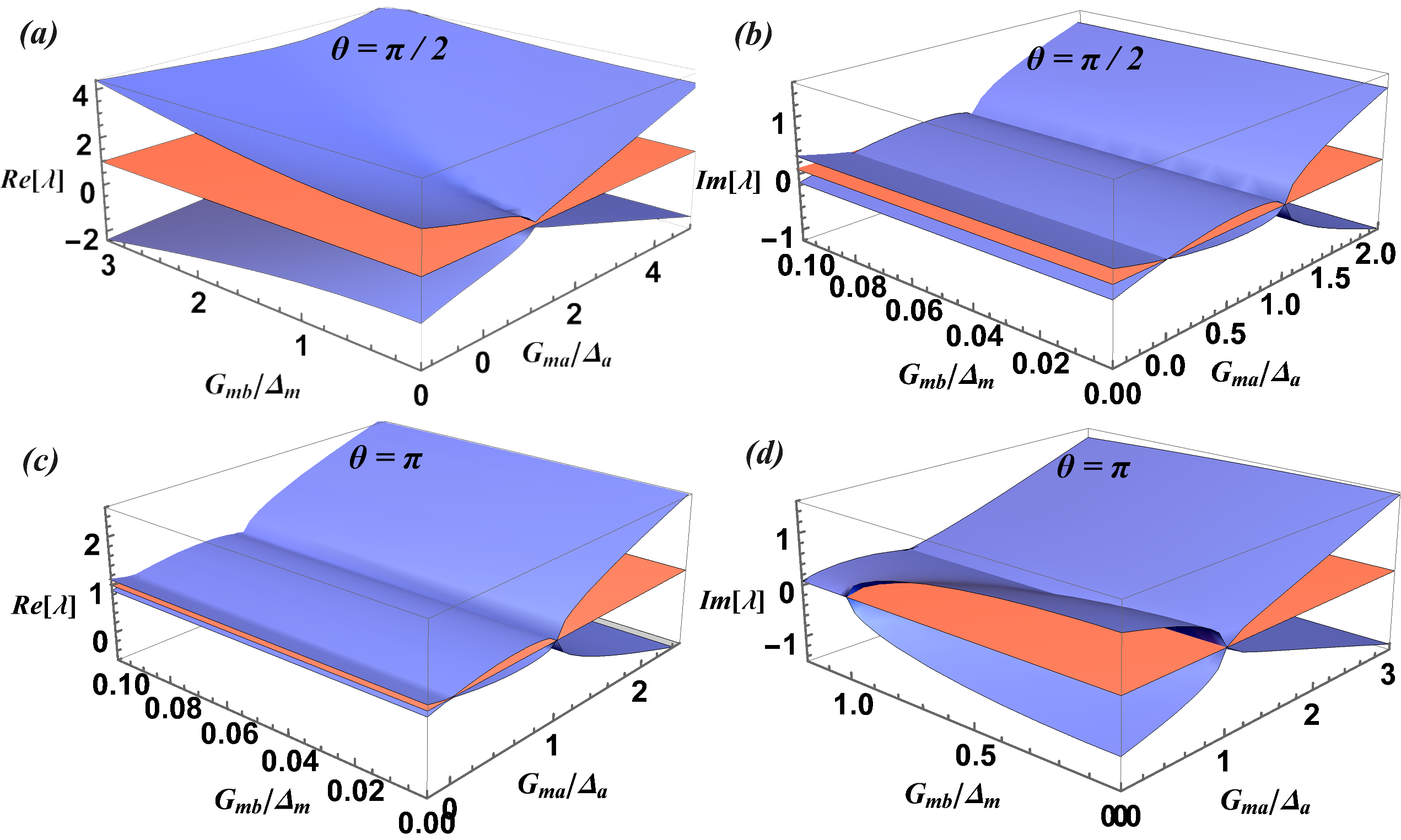}
	\caption {Eigenvalue spectrum, real $Re[\lambda]$ (left column) and imaginary part $Im[\lambda]]$ (right column), as a function of normalized magnon-photon coupling $ G_{ma}/\Delta_a $ and normalized magnon-phonon coupling $ G_{mb}/\Delta_b $, at $ \Gamma/\omega_b=1 $. For (a) and (b), the angle of traveling field is $\theta=\pi/2$, while for (a) and (b) angle is $\theta=\pi/2$. The other parameters are the same as in Fig.2 of the main text.} 
	\label{figs2}
\end{figure}

Figure \ref{figs1} illustrates the 3D illustration of the system's energy eigenvalues as functions of the magnon-photon coupling $ G_{ma}/\Delta_a $ and magnon-phonon coupling $ G_{mb}/\Delta_b $. Figs.\ref{figs2}(a), \ref{figs2}(b) show the real and imaginary parts of the eigenvalues when the incident angle of the traveling field is $ \pi/2 $. It can be observed that when the photon-magnon coupling is zero or equal to the $ G_{mb}/\Delta_b $, the imaginary parts of the system's eigenvalues are zero, indicating real eigenvalues. Notably, a third-order EP appears only when the magnon-phonon coupling rate is extremely low. As the magnon-phonon coupling rate increases, no merging of the eigenvalues occurs, and the system tends towards instability, as confirmed by the stability analysis discussed in main text.

For the incident angle $ \pi $, shown in Figs.\ref{figs2}(c), \ref{figs2}(d), the third-order EP also appears on the line where the magnon-photon coupling is equal the magnon-phonon coupling. In this scenario, the system is protected by PT symmetry and has real eigenvalues. Similarly, the third-order EP is present only when the magnon-phonon coupling rate is low. As this coupling rate increases, the system's eigenvalues do not merge, and the system becomes unstable.